\documentclass[twocolumn,aps,prl,superscriptaddress,groupedaddress]{revtex4}
\usepackage{textcomp}
\usepackage{amssymb}
\usepackage{amsmath,amssymb}
\usepackage{dsfont}
\usepackage{graphicx,epstopdf}
\DeclareMathOperator{\Tr}{Tr}
\begin{document}

\title{Time-bin to Polarization Conversion of Ultrafast Photonic Qubits}
\
\author{Connor~Kupchak}
\affiliation{Department of Physics, University of Ottawa, Ottawa, Ontario, K1N 6N5, Canada}
\affiliation{National Research Council of Canada, 100 Sussex Drive, Ottawa, Ontario, K1A 0R6, Canada}
\author{Philip~J.~Bustard}
\author{Khabat~Heshami}
\affiliation{National Research Council of Canada, 100 Sussex Drive, Ottawa, Ontario, K1A 0R6, Canada}
\author{Jennifer~Erskine}
\affiliation{Department of Physics, University of Ottawa, Ottawa, Ontario, K1N 6N5, Canada}
\affiliation{National Research Council of Canada, 100 Sussex Drive, Ottawa, Ontario, K1A 0R6, Canada}
\author{Michael~Spanner}
\author{Duncan~G.~England}
\affiliation{National Research Council of Canada, 100 Sussex Drive, Ottawa, Ontario, K1A 0R6, Canada}
\author{Benjamin~J.~Sussman}
\affiliation{Department of Physics, University of Ottawa, Ottawa, Ontario, K1N 6N5, Canada}
\affiliation{National Research Council of Canada, 100 Sussex Drive, Ottawa, Ontario, K1A 0R6, Canada}

\begin{abstract}
The encoding of quantum information in photonic time-bin qubits is apt for long distance quantum communication schemes. In practice, due to technical constraints such as detector response time, or the speed with which co-polarized time-bins can be switched, other encodings, e.g. polarization, are often preferred for operations like state detection.  Here, we present the conversion of qubits between polarization and time-bin encodings using a method that is based on an ultrafast optical Kerr shutter and attain efficiencies of 97\% and an average fidelity of 0.827$\pm$0.003 with shutter speeds near 1 ps.  Our demonstration delineates an essential requirement for the development of hybrid and high-rate optical quantum networks. 
\end{abstract}

\maketitle

The encoding of quantum information (QI) into photons holds much promise in numerous future technologies.  The QI can be mapped onto various degrees of freedom that are used as basis-states.  One attractive option is to encode onto qubits composed of two co-polarized but temporally distinct wave packets, or time-bins; these basis states are often labelled by their arrival time as \emph{early} ($|e\rangle$) and \emph{late} ($|l\rangle$)~\cite{Gisin1999}.  Time-bin encodings have recently been used in the successful transmission of qubits over hundreds of kilometers~\cite{Taksue2013}, and in teleportation using real-world fiber networks~\cite{Valivarthi2016, Sun2016}.  

The disadvantage is that direct readout of information encoded in time-bins can require the peak-to-peak separation $\Delta\tau_{el}$ between $|e\rangle$ and $|l\rangle$ to be sufficiently greater than the response time of the detector, and can impose a minimum time for the bin separation.  Typical detector response times correspond to a bin separation of at least nanoseconds~\cite{Valivarthi2014}; this limits the available bandwidth for encoding and can necessitate active interferometric stabilization when preparing and detecting qubits and qudits~\cite{Ikuta2017}.  Recently, advanced methods have emerged that utilize nonlinear techniques to creatively detect light states encoded in temporal modes~\cite{Donohue2013, Ansari2017}, however these implementations are constrained to operate at low efficiencies. 

\begin{figure*}[htb]
\includegraphics[width=2\columnwidth]{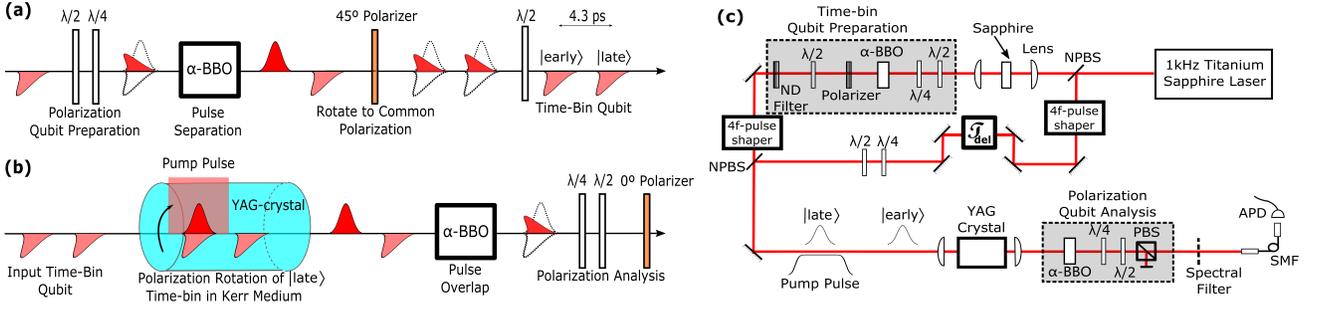} \protect\protect\caption{ (a) Schematic diagram for time-bin qubit preparation starting with a polarization state and (b) conversion of time-bin qubits to the polarization degree of freedom using the OKS; (c) corresponding experimental setup.}
\end{figure*}

Polarization encoding is a popular choice for various QI applications \cite{BB84, Bouwmeester1997, Kok2007} but can be problematic for long distance implementations~\cite{Bonato2006, Versteegh2015, Antonelli2011}.  Ideally one would have the flexibility to convert arbitrary photonic states between encodings depending on the application, e.g. a time-bin encoding for transmission and a polarization encoding for state detection and manipulation.  Many previous schemes for time-bin to polarization qubit conversion are lossy and rely on post-selection using passive optics~\cite{Sanaka2002, Takesue2005, Bussieres2010, Tanzilli2013}.   Another approach could rely on active switches involving Pockels cells or electro-optical modulation to convert between encodings by rotating the polarization state of a single bin~\cite{Soudagar2007}.  For these active implementations, the rise time of the device limits the temporal separation of the bins and restricts the data transfer rate.  Typical switching devices that sell commercially, have rise times on the order of nanoseconds, and shorter times of $<$~100~ps are achievable in non-commercial waveguides~\cite{Melikyan2014}, but usually exhibit insertion losses of 1-3 dB.  Recently, all-optical solutions based on cross-phase modulation (XPM) for converting time-bins between encodings have been developed that can switch as fast as 50~ps \cite{Nowierski2016}. It is therefore desirable to progress these all-optical conversion methods to higher bandwidths and operational speeds in the ultrafast regime.

Here, we realize an efficient scheme for the conversion of qubits between time-bin and polarization encodings, and demonstrate its potential using ultrafast laser pulses attenuated to the single-photon level.  Our approach is reversible, and capable of bandwidths greater than 200~GHz.   Devices of this functionality may find use in high-bandwidth quantum communication networks and enable the interfacing of time-bin qubits with ultrafast quantum memories  \cite{England2015} in local QI processing.

Our scheme is based on the optical Kerr effect: induced birefringence in a $\chi^{(3)}$ nonlinear medium which is proportional to the irradiance of an applied pump field.  We use this effect to map photons between polarization states.  Typically, a $\chi^{(3)}$ medium is placed between two axis-crossed polarizers so that the input probe light is blocked except in the presence of an applied pump field;  such a setup is referred to as an optical Kerr shutter (OKS)~\cite{Orszag1980, LaGasse1989, Yu2003}.  The shutter efficiency $\eta$ is given by~\cite{Kanbara1994}

\begin{equation}
\eta=\sin^2\left( 2\theta \right) \sin^2\left(\frac{\Delta\phi}{2}\right),
\end{equation}
where
\begin{equation}
\Delta\phi=\frac{2\pi n_2 L_{\text{eff}}I}{\lambda_{\text{probe}}}
\end{equation}
is the phase shift induced by the pump field of intensity $I$, $n_2$ is the nonlinear component of the refractive index, $L_{\text{eff}}$ is the effective length of the medium, $\lambda_{\text{probe}}$ is the wavelength of the probe field, and $\theta$ is the polarization angle between the pump and probe field.  The case of $\Delta\phi=\pi$ corresponds to the probe field undergoing a full 90$^{\circ}$ polarization rotation i.e. horizontal flipped to vertical.

Our experimental scheme for the qubit conversion process can be divided into two main parts displayed in Figs.~1(a)-(b). First, the time-bin qubit preparation stage shown in Fig.~1(a) where qubits are initially encoded into polarization states using a half-waveplate (HWP) and quarter-waveplate (QWP) combination.  The polarization qubits then enter a birefringent medium to temporally separate the horizontal and vertical components; establishing the $|e\rangle$ and $|l\rangle$ time bins, respectively.  In our setup, a 10~mm long $\alpha$-BBO crystal sets the separation between the $|e\rangle$ and $|l\rangle$ time-bins to be $\Delta\tau_{el}=$4.3~ps.  After this stage, the qubits pass through a polarizer set to transmit diagonal linear polarization~(45$^{\circ}$) resulting in a 50\% loss and an additional HWP to prepare time-bins that are horizontally co-polarized.  

The second part, shown in Fig.~1(b), is the procedure for converting time-bins to a polarization encoding.  To do so, we temporally overlap the pump field with the $|l\rangle$ time-bin and focus both fields into the Kerr medium to rotate the polarization of the $|l\rangle$ bin from a horizontal to vertical polarization due to the OKS operation.  This is followed by transmission through a second, identical $\alpha$-BBO crystal with its axis rotated by 90$^{\circ}$ with respect to the first, such that the now orthogonally polarized time bins are overlapped into a single temporal bin.  This completes the mapping of the qubits to a polarization-based encoding that is suited for measurement and manipulation by common polarization state analysis techniques.  Note that the OKS could also be implemented to perform the reverse operation i.e. from a polarization to time-bin encoding.  In this case, a polarization qubit is first sent through a birefringent material to achieve temporal mode separation of the polarization states followed by an OKS operation on the $|l\rangle$ time-bin. 

A diagram of our experimental setup is given in Fig.~1(c).  The pump beam is derived from a 1~kHz repetition rate, chirped pulsed amplifier laser emitting pulses with a 90~fs duration at a wavelength of 800~nm.  The probe field is generated by splitting-off a portion of the original pump pulse for use in a white light source generated in sapphire \cite{Bradler2009}.  Before collinear combination on a non-polarizing beam splitter (NPBS), both pump and probe beams are spectrally tailored using independent 4f-shapers constructed with adjustable razor blades at the focal plane to serve as a mask \cite{Weiner2011}. The probe beam is set to a central wavelength of 710 nm and bandwidth of $\Delta\lambda_{\text{probe}}=$5.7~nm with a full width at half maximum (FWHM) duration of $\sim$270~fs while a narrowband pump beam is created by filtering to a top-hat-shaped spectrum of $\Delta\lambda_{\text{pump}}=$1.8~nm and a pulse duration of $\Delta\tau_{\text{pump}}\sim$~1.15~ps.  The difference in bandwidths is necessary to achieve a quasi-uniform pump intensity over the probe duration.

For our Kerr medium we employ a $\langle111\rangle$-cut, 8~mm long yttrium aluminum garnet (YAG) crystal of no inherent birefringence, chosen for its relatively high $n_2$ value~\cite{Adair1989}.  The probe field is spatially filtered to achieve a Gaussian spatial mode with a beam waist in the focal plane of 20~$\mu$m compared to 60~$\mu$m for the pump.  A set of waveplates and a polarizing beam splitter~(PBS) are situated after the second $\alpha$-BBO crystal for polarization state projection; this is succeeded by a series of spectral filters to extinguish the pump field and permit measurement at the single-photon level via coupling to an avalanche photodiode~(APD) using a single-mode fiber~(SMF). 

{\it Single time-bin OKS operation.-} First, we characterize the efficiency of our OKS operation using attenuated pulses defined in a single temporal bin.  Here, we set the energy of the pump pulse to 840~nJ and fix the polarization to 45$^{\circ}$ (diagonal) with respect to the horizontally polarized time-bin.  To investigate single-photon-level conditions, the mean photon number $\langle n\rangle$ of our probe pulse is set at 1.17$\pm$0.12 and counts are recorded on the APD gated with a 2~ns window at the 1~kHz repetition rate of the main laser.  By scanning the temporal delay of the pump pulse $\tau_{\text{del}}$ with respect to the probe and setting the analysis optics to transmit a vertical polarization, we identify the peak shutter efficiency~(Fig.~2(a)).

\begin{figure}[htb]
\includegraphics[width=1\columnwidth]{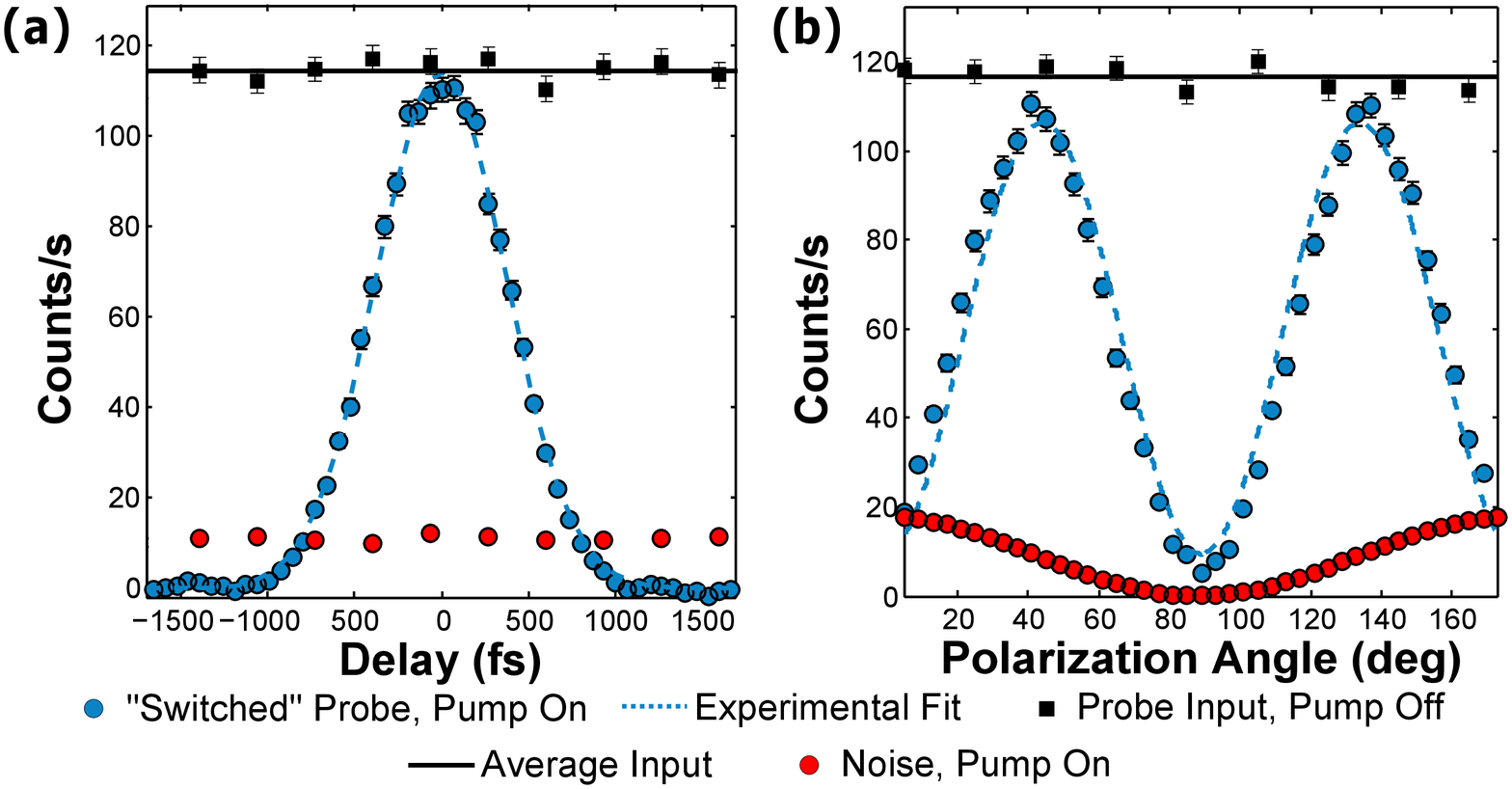}\protect\protect\caption{The OKS operating on a single time bin with (a) counts per second measured as the pump pulse delay is scanned.  The counts corresponding to the transmitted, polarization-rotated, probe pulse (blue circles) with fit (blue dashed line) are compared to the original, horizontally polarized input pulses (black squares) and their mean (solid line) and the noise counts (red circles).  (b) Dependence of the OKS switching on the relative polarization angle $\theta$, between the pump and probe pulses and corresponding fit. The error bars are derived from Poisson statistics.}
\end{figure}

Under these conditions, we observe near-perfect polarization rotation of the probe pulse from horizontal to vertical with a peak shutter efficiency of $\eta=0.97\pm0.03$ relative to the original input pulse.  The efficiency is evaluated using the peak counts in Fig.~2(a) to represent the maximum OKS related signal counts $N_{\text{OKS}}$, the noise counts due to the pump beam $N_{\text{noise}}$, and the counts corresponding to the original input pulse $N_{\text{input}}$ such that $\eta=(N_{\text{OKS}}-N_{\text{noise}})/N_{\text{input}}$.  It is also important to attain a sufficiently high signal-to-noise-ratio (SNR) in order to distinguish the state of the qubit.  A SNR of $9.2\pm0.3$ is achieved for the OKS signal fit relative to $N_{\text{noise}}$; this value is comparable to other quantum channels designed for time-bin qubits~\cite{Gundogan2015}.  From the fit in Fig. 2(a), we can also evaluate the operating speed of the OKS and find a FWHM of $\Delta\tau_{\text{OKS}}=0.88\pm0.01$~ps.  Combined with $\Delta\tau_{el}$, this establishes a potential bandwidth of our device of over 200~GHz when operating on THz-bandwidth photons. Note that the $\sin^2(\Delta\phi/2)$ response of the OKS (see Eq. (1)) yields a FWHM that is less than the pump pulse duration~\cite{Dmitruk2016}.

In order to verify the OKS operation with respect to the polarization angle $\theta$, the pump polarization is rotated over a range of 180$^{\circ}$ and the corresponding polarization-rotated probe pulses are collected on the APD.  Here, the pump pulse energy remains at 840~nJ and the temporal delay between the pulses is fixed to zero.  As can be seen in Fig.~2(b),  along with the expected $\sin^2(2\theta)$ behavior in accordance with Eq.~(2), the noise counts follow the polarization of the pump field.%

Lastly, we characterize the performance of the OKS as a function of the energy of the pump pulse (Fig.~3).  For this analysis, we fix the pump delay and polarization to the optimal values and measure the OKS efficiency and noise.  From Fig. 3 it is clear that a range of optimal pump energies emerge between 800-900~nJ where the OKS efficiency approaches 100\% and the noise photon rate remains low enough to yield a SNR of $\sim$ 10.  At energies greater than this range we observe a sharp, nonlinear increase in noise photons that can be attributed to spectral broadening of the pump pulse by self-phase modulation (SPM) in the YAG crystal.

\begin{figure}[htb]
\includegraphics[width=1\columnwidth]{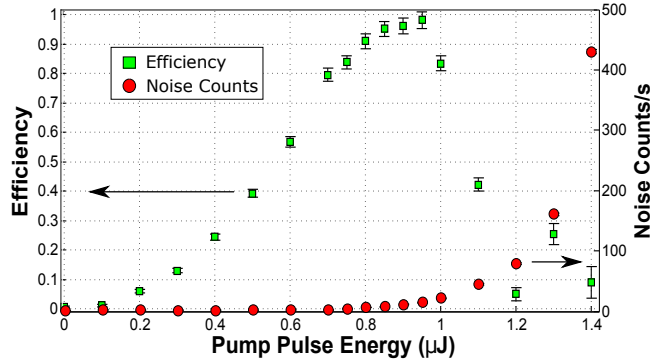}
\caption{OKS efficiency (left ordinate: green squares) and noise counts per second (right ordinate:  red circles) with respect to the energy of the pump pulse. Error bars on the noise counts are $\sim\pm1$ and smaller than the dots shown. }
\end{figure}

\begin{figure*}[htb]
\includegraphics[width=2\columnwidth]{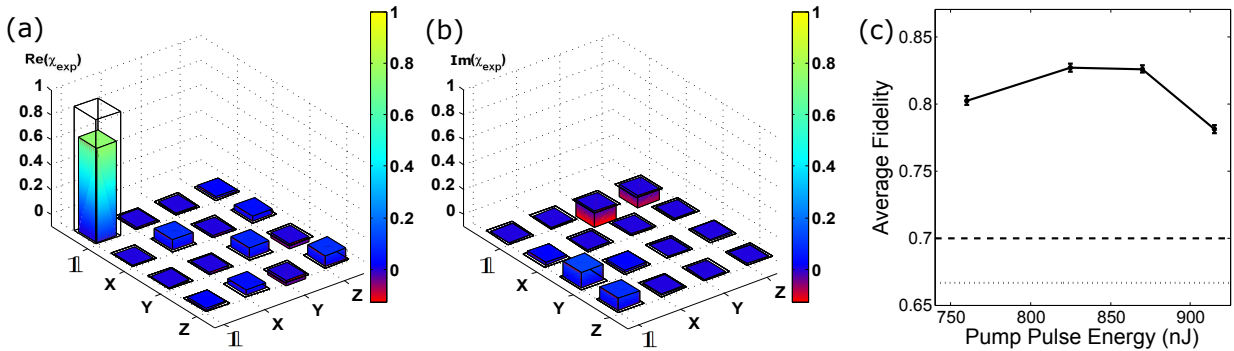}
\protect\protect\caption{(a) Real and (b) imaginary  components of the experimentally reconstructed process tensor (filled bars) in the basis of Pauli matrices as compared to the ideal process tensor values (wire-grid). On-diagonal elements represent components for each of the Pauli operators $\{\mathds{1}, X, Y, Z\}$. (c) Behavior of the average fidelity (black circles) at pump energies near the optimum compared to the classical thresholds; this is 2/3 for the case of single photons (dotted line) and 0.70~(dashed line) for the mean photon number and efficiency used in our study.}
\end{figure*}

{\it Time-bin to polarization qubit conversion.-} With the OKS operation characterized, we turn our attention to mapping time-bin qubits to a polarization encoding according to the scheme depicted in Fig.~1.  The input time-bin qubits  are prepared with a mean photon number of $\langle n\rangle=0.75\pm0.06$ and the pump pulse energy set to 825~nJ.  A mean photon number of $\sim$0.7 is consistent with typical heralding efficiencies for spontaneous parametric down-conversion single photon sources \cite{Ramelow2013}.  We therefore expect that the fidelity observed here reflects what could be achieved using heralded single photons.  In our implementation, the process-related losses stemming from the YAG sample correspond to an optical transmission of 82\% and contributes to the overall channel efficiency of 11$\pm$1\% when also including the fiber coupling efficiency, APD response and transmission losses.

To quantify the performance of our qubit conversion scheme we perform quantum process tomography~\cite{Nielsen1997} on the converted time-bin qubits.  Process tomography is accomplished by using our preparation waveplates to generate 6 input polarization states ($|H\rangle$, $|V\rangle$, $|D\rangle=\frac{1}{\sqrt2}(|H\rangle+|V\rangle)$, $|A\rangle=\frac{1}{\sqrt2}(|H\rangle-|V\rangle)$, $|R\rangle=\frac{1}{\sqrt2}(|H\rangle +i |V\rangle)$, $|L\rangle=\frac{1}{\sqrt2}(|H\rangle -i |V\rangle)$ where $H$ indicates horizontal polarization and $V$ vertical polarization;  these form three mutually unbiased bases in the qubit Hilbert space.  These six states are first converted to their corresponding time-bin counterparts (i.e. $|H\rangle$ goes to $|e\rangle$ and $|V\rangle$ goes to $|l\rangle$) as shown in Fig.~1(a).  Upon conversion back to the polarization degree of freedom by the OKS scheme in Fig.~1(b),  the output state is projected onto all six polarization states using the analyzer waveplates.  This 36-element data set forms an over-complete basis and allows us to experimentally reconstruct the process tensor $\chi_{\text{exp}}$ in the Pauli operator basis $\sigma_{\text{i=1..4}}\equiv\{\mathds{1}, X, Y, Z\}$, similar to previous work \cite{Silverstone2015}.  Proper conversion corresponds to an identity operator that is defined by unity at the $(\mathds{1},\mathds{1})$ element and zeros otherwise. 

Fig.~4(a-b) shows the real and imaginary components of the experimentally-reconstructed process matrix for the qubit conversion, where elements of $\chi_{\text{exp}}$ determine a completely positive map $\mathcal{E}(\rho_{\text{in}})=\sum\chi_{\text{exp}_{ij}}\sigma_i\rho_{\text{in}}\sigma_j=\rho_{\text{out}}$ that characterizes the quantum channel.  The fidelity of the reconstructed process matrix $\chi_{\text{exp}}$ compared to the ideal case $\chi_{\mathds{1}}$ is calculated by $F_{\text{proc}}(\chi_{\mathds{1}},\chi_{\text{exp}})=\left(\Tr\sqrt{\sqrt{\chi_{\mathds{1}}}\chi_{\text{exp}}\sqrt{\chi_{\mathds{1}}}}\right)^2$ to produce a process fidelity of 0.740$\pm$0.005.  The uncertainty is estimated by including Poissonian noise to the recorded counts; the corresponding error bars represent the FWHM of the distribution. The average fidelity is calculated via $F_{\text{avg}}=(2F_{\text{proc}}+1)/3$ \cite{Nielsen2002} to yield $F_{\text{avg}}=$0.827$\pm$0.003.  Given our efficiencies and the mean photon number used, this value exceeds the average fidelity of 0.70 (shown in Fig.~4(c)) needed for our device to faithfully operate as a quantum channel \cite{Gundogan2012} and verifies that the qubit states are indeed preserved when converting from time-bin to polarization encodings.

Finally, we investigate how the average fidelity behaves as a function of pump energy. The corresponding values are summarized in Fig.~4(c) where we can see reductions in the fidelity on either side the optimum at 825~nJ.  The lower pump energies correspond to lower conversion efficiencies on the $|l\rangle$ bin and leads to an unintended ratio of horizontal and vertical components.  As a result, the combined temporal mode contains an improper polarization when projected onto a measurement basis.  Furthermore, the remaining, non-polarization rotated photons in the $|l\rangle$ bin can also be erroneously recorded on the APD and reduce the visibility between orthogonal states.  At the higher pump energies, the increase in self-phase modulation-related~\cite{Bradler2009} noise reduces the ability to correctly discriminate the polarization state, thereby decreasing the SNR and fidelity.  In future implementations, the fidelity of the process could be increased by using shorter probe wavelengths due to the reduction in the pump energy needed to achieve $\Delta\phi=\pi$.  Here, the SNR would likewise increase with spectral separation between pump and probe fields due to fewer SPM noise photons created at the probe wavelength.  Overall improvements to our scheme are possible by using anti-reflective coatings on the faces of the YAG sample and by decreasing the pump power required for conversion by moving to a fiber system. 

In summary, we present a platform for ultrafast polarization rotation that enables conversion of qubits between time-bin and polarization encodings. The technique is reversible, highly efficient, and leaves the spectrum of the photon unchanged and thus adds a valuable tool to the suite of ultrafast protocols designed to measure time-bin qubits~\cite{Donohue2013, Ansari2017}.  Our switch operates at picosecond timescales to allow time-bin encodings that are orders of magnitude faster than typical detector response times ($>100$ ps) and permits high-bandwidth quantum communication without requiring complex stabilized interferometers~\cite{Li2016}.  In addition to communications, we expect our scheme to find applications in photonic quantum information processing, such as linear quantum computing in a single spatial mode~\cite{Humphreys2013} and to offer a path towards architectures with hybridized encodings and higher dimensional quantum states that can benefit from efficient and ultrafast operations. Beyond quantum optics, our OKS properties could be applied to areas where efficient, ultrafast switching of weak signals at low noise would be of benefit, for example time-resolved spectroscopy~\cite{Takeda2000, Dmitruk2016} or nonlinear microscopy~\cite{Matousek1999, Blake2016}.  Our study of the noise processes in the OKS at the single photon level provide a benchmark for these applications.  Implementation of our approach in a waveguide will enable low-power operation and integration into more compact setups for a range of photonic applications.

\section*{Funding Information}
This work is supported by the Natural Sciences and Engineering Research Council of Canada.

\section*{Acknowledgments}

The authors wish to thank Paul Hockett, Rune Lausten, Kent Fisher and Neil Sinclair for helpful discussions and comments.

\end{document}